\documentclass[useAMS,usenatbib,usegraphicx,letter]{mn2e}
\usepackage{times}
\usepackage{aas_macros}

\title[Infrared interferometry of CH Cyg]{Detection of non--radial pulsation and faint companion in the symbiotic star CH Cyg}

\author[E.  Pedretti et  al.]   {E.~Pedretti$^1$\thanks{Affiliated to Scottish universities physics alliance (SUPA).},  J.~D.~Monnier$^2$,  S.~Lacour$^3$, W.~A.~Traub$^4$, W.~C.~Danchi$^5$, P.~G.~Tuthill$^6$,
\newauthor 
N.~D.~Thureau$^1$, R.~Millan--Gabet$^7$,  J--P.~Berger$^3$, M.~G.~Lacasse$^8$, P.~A.~Schuller$^9$ 
\newauthor 
F.~P.~Schloerb$^{10}$ and N.~P.~Carleton$^8$ \\
$^1$School of  Physics and Astronomy, University of  St Andrews, North Haugh, St Andrews KY16 9SS, Scotland.\\ 
$^2$University of Michigan, Astronomy dept., 914 Dennison bldg., 500 Church street, Ann Arbor, MI, 40109, USA\\
$^3$Laboratoire d'Astrophysique de l'Observatoire de Grenoble (LAOG), 414 rue de la Piscine, BP 53--X, Grenoble, France.\\
$^4$Jet Propulsion Laboratory, { California Institute of Technology}, M/S 301--451, 4800 Oak Grove Drive, Pasadena, CA, 91109, USA.\\
$^5$NASA Goddard Space Flight Center, 8800, Greenbelt Road, Greenbelt, MD20771, USA\\
$^6$School of Physics, Sydney University, N. S. W. 2006, Australia\\
$^7$Michelson Science Center, California Institute of Technology, 770 S. Wilson Ave. MS 100--22, Pasadena, CA 91125, USA.\\
$^8$Harvard--Smithsonian  Center for Astrophysics, 60 Garden Street, Cambridge, MA, 02138, USA.\\
$^9$Institut d'Astrophysique Spatiale, Université Paris--Sud, b\^atiment 121, 91405, Orsay Cedex, France.\\
$^{10}$University of Massachusetts, Astronomy Department, Amherst, MA, 01003--4610, USA.
} 

\pagerange{\pageref{firstpage}--\pageref{lastpage}} \pubyear{2008}

\def\LaTeX{L\kern-.36em\raise.3ex\hbox{a}\kern-.15em
    T\kern-.1667em\lower.7ex\hbox{E}\kern-.125emX}

\def\deg{$^\circ$}

\def\sun{$_\odot$~}
\def\uv{$uv$~}
\def\abt{$\sim$}
\newcommand\textsupsub[2]{$^{#1} _{#2}$}

\begin{document}
\label{firstpage}

\maketitle
\begin{abstract} We have detected asymmetry in the symbiotic star CH~Cyg through the measurement of precision closure--phase with the IONIC beam combiner, at the IOTA interferometer. The position of the asymmetry changes with time and is correlated with the phase of the 2.1--yr period found in the radial velocity measurements for this star. We can model the time--dependent asymmetry either as the orbit of a low--mass companion around the M~giant or as an asymmetric, 20\% change in brightness across the M~giant. We do not detect a change in the size of the star during a 3 year monitoring period neither with respect to time nor with respect to wavelength. We find a spherical dust--shell with an emission size of { 2.2$\pm$0.1~D$_*$}~FWHM around the M~giant star{. The star to dust flux ratio is estimated to be} 11.63$\pm$0.3. While the most likely explanation for the 20\% change in brightness is non--radial pulsation we argue that a low--mass companion in close orbit could be the physical cause of the pulsation. The combined effect of pulsation and low--mass companion could explain the behaviour { revealed by} the radial--velocity curves and the time--dependent asymmetry detected in { the} closure--phase data. If CH~Cyg is a typical long secondary period variable then { these} variations could be explained by the effect of  an orbiting low--mass companion on the primary star.

\end{abstract}

\section{Introduction}
\begin{table*}
\centering
\caption{\label{observation_log} Log of observations. The IOTA 3T telescope configuration refers to the location of the A, B, C telescopes along the NE, SE and NE arms.}
\begin{minipage}{140mm}
\begin{tabular}{l c c c c c c r}
\hline
\hline
{Date}& {Mean}& {Phase}& {Telescope}& {${\lambda}$}& {${\Delta}$${\lambda}$ }& R\footnote{Only applicable to the IOTA spectrograph.}  &{Calibrator}\\
{(UT)}& {JD}&    &     & {(${\mu}$m)}& {(${\mu}$m)}& &{names}\\
\hline
{2004Apr23}& {2453119}& {0.40}& {IOTA , A35{}--B15{}--C10}& {1.51}& {0.090}& & {${\alpha}$ Lyr, ${\alpha}$ Aql}\\
{2004Apr24}& {2453120}& {0.40}& {IOTA , A35{}--B15{}--C10}& {1.64}& {0.100}& & {${\alpha}$ Lyr, ${\alpha}$ Aql}\\
{2004Apr25}& {2453121}& {0.40}& {IOTA , A35{}--B15{}--C10}& {1.64}& {0.100}& & {${\alpha}$ Lyr, ${\alpha}$ Aql}\\
{2004Apr26}& {2453122}& {0.40}& {IOTA , A35{}--B15{}--C10}& {1.78}& {0.090}& & {${\alpha}$ Lyr, ${\alpha}$ Aql}\\
{2004Apr29}& {2453125}& {0.41}& {IOTA , A35{}--B15{}--C10}& {1.78}& {0.090}& & {${\alpha}$ Lyr, ${\rho}$ Ser}\\
{2004Apr30}& {2453126}& {0.41}& {IOTA , A35{}--B15{}--C10}& {1.78}& {0.090}& & {${\alpha}$ Lyr, ${\nu}$ Hya}\\
{2004May01}& {2453127}& {0.41}& {IOTA , A35{}--B15{}--C10}& {1.78}& {0.090}& & {${\alpha}$ Lyr, ${\alpha}$ Aql}\\
{2004Sep04}& {2453105}& {0.38}& {Keck A, Golay mask}& {1.64}& {0.025}& & {${\alpha}$ Lyr}\\
{2005Jun06}& {2453528}& {0.94}& {IOTA , A35{}--B15{}--C10}& {1.66}& {0.300}& & {${\alpha}$ Lyr}\\
{2005Jun08}& {\raggedleft2453530\par}& {0.95}& {IOTA , A25{}--B15{}--C10}& {1.66}& {0.300}& & {${\alpha}$ Lyr}\\
{2006Apr24}& {2453850}& {0.37}& {IOTA , A35{}--B15{}--C10}& {1.66}& {0.300}& 39& {${\alpha}$ Lyr, ${\beta}$Her}\\
{2006Apr30}& {\raggedleft2453856\par}& {0.38}& {IOTA , A35{}--B15{}--C10}& {1.66}& {0.300}&39 & {${\alpha}$ Lyr, ${\beta}$Her}\\
{2006May01}& {2453857}& {0.38}& {IOTA , A35{}--B15{}--C10}& {1.66}& {0.300}& 39& {${\alpha}$ Lyr, ${\beta}$Her}\\
{2006May02}& {\raggedleft2453858 \par}& {0.38}& {IOTA , A35{}--B15{}--C10}& {1.66}& {0.300}&39 & {${\alpha}$ Lyr, ${\beta}$Her}\\
\hline
\end{tabular}
\end{minipage}
\end{table*}

Symbiotic stars are objects presenting combination spectra of a hot ionised nebula and the cool continuum absorption molecular features of a late--type star. Nowadays, symbiotic stars are understood as mass--transfer binaries of short period, from a few to 10 years. The separation can vary from a few AU to slightly more than 10 AU. The symbiotic pair is usually composed of a cool giant star with an accreting compact object, either a white dwarf or a neutron star.

CH Cyg is one of the most studied of symbiotic variables. The star presents a composite spectrum of a M6--7 giant star during quiescent phase and a hot component blue continuum from 6000 to 9000~K temperature and low excitation line spectrum during the active phase \citep{1974PASP...86..233D}. \cite{1975MNRAS.171..171W} classified the star as an S--type symbiotic with no hot dust, but long term multi--wavelength  photometry study of  the star \citep{1988Ap&SS.146...33T} has shown that hot dust appeared in the system after the 1984 outburst. The dust was modelled as a spherical shell of inner radius of 15 AU by \cite{2001ARep...45..797B} { through spectral energy distribution (SED) fitting}.

\cite{1998AJ....116..981D} measured an angular diameter of 10.4~mas at 2.2~$\mu$m for CH~Cyg with infrared interferometry. \cite{2000SPIE.4006..472Y} observed CH~Cyg with the Cambridge optical aperture synthesis telescope (COAST) in 1999. The obtained visibility and closure--phase data were best modelled by an elliptical, limb--darkened star. Their interpretation of these findings was that the ellipticity of the star was either due to an extension of the  M~giant atmosphere or to the partial eclipse of an orbiting red giant companion as proposed by \cite{1996A&A...308L...9S}. 

CH~Cyg shows photometric and radial--velocity variations. In photometry two main periods are found: a small amplitude (0.1 magnitude) \abt100 day period \citep{1992A&A...254..127M} likely caused by stellar pulsation and a  \abt770 day period \citep{1992A&A...254..127M} or a \abt1 magnitude, \abt750 day period \citep{2007AN....328..909S}. These photometric variations are not always detected \citep{1996A&A...311..484M} { and are not related to the 100--day pulsation period}. \cite{1993AJ....105.1074H} pointed out that the photometric variations of CH~Cyg are far longer than the fundamental pulsation mode for this star{, which is a first--overtone pulsator \citep{1992A&A...254..127M}}. { The radial--velocity variations from the literature \citep{1993AJ....105.1074H}} show two periods: a 15.6--yr long period and a 2.1--yr (750 days) short period.

\cite{1993AJ....105.1074H,2008arXiv0811.0631H} suggested a correlation between the 750/770 days photometric period and the 750 days radial--velocity period. They also remarked that the photometric variations of CH~Cyg are similar to those found in long secondary period (LSP) variable stars \citep{2006MmSAI..77..523H}. This type of variability is found in some semi-regular variables and in about the 25\% { of the} Large Magellanic Cloud (LMC) semi-regular variables. \cite{2002AJ....123.1002H} and \cite{2004ApJ...604..800W}  found that several semi-regular variables also show spectroscopic { behaviour} consistent with LSP variability.  The cause of { the} LSP is currently unknown but possible explanations are highlighted in \cite{2004ApJ...604..800W}. They conclude that the most likely explanation for LSP variations is a low order non radial pulsation on the outer radiative layers of the giant star.

\cite{1999IAUS..191..151W} also discovered that LSP variables follow a period--luminosity (P--L) relation which he called ``sequence~D''. \cite{2004AcA....54..347S}  noted that Wood sequence--D variables overlap with sequence--E contact binaries, implying that sequence--D is indeed a class of binaries. \cite{2007ApJ...660.1486S} also found that 5\% of LSPs in the LMC present ellipsoidal--like or eclipsing--like modulation that are usually shifted in phase with respect to LSP light curves. 

\cite{1993AJ....105.1074H} proposed a model where CH~Cyg was a triple system with the symbiotic pair in a 2.1--yr orbit. The reasons for having the symbiotic pair on the 2.1--yr orbit were that no known S--type symbiotic star had orbital period larger than 5 years, the 2.1--yr period was too long for a M~giant fundamental--mode pulsation and there was weak evidence for a high { inclination} 15.6--yr orbit.  The third star was either regarded as  a  G--K  dwarf  \citep{1993AJ....105.1074H}  or  a  M~giant \citep{1996A&A...308L...9S}. The inclination of the 15.6--yr orbit was unknown at the time but was recently inferred from the several eclipses reported in the literature \citep{1987Ap&SS.131..733M,2002MNRAS.335..526E,2003ApJ...584.1021S}. \cite{2006A&A...446..603S} suggested that the 2.1--yr period was caused by a pulsation in the M~giant and not by a close binary. 

There is controversy on the shape of the possible orbit of the close pair. \cite{1993AJ....105.1074H} argued that the asymmetric line profiles could be caused by a M~giant star irradiated by a white dwarf. An asymmetric line profile could lead to a false elliptic solution for an orbit obtained from radial velocities. According to \cite{1993AJ....105.1074H} the orbit of CH Cyg should be circular due to tidal interaction with the M~giant.

\cite{2008arXiv0811.0631H}  re--examined the conclusions of the \cite{1993AJ....105.1074H} paper. They concluded that the 2.1--yr velocity variation is consistent with LSP variation and that the white dwarf responsible for the activity in the system is on the 15.6--yr orbit. The 2.1--yr period would be caused either by non--radial pulsation of the star or by a low--mass companion in close orbit to the M~giant.

This paper  presents the results of infrared interferometric observations performed in 2004--2006 at the infrared optical telescope array (IOTA) \citep{2004SPIE.5491..482T} and at the Keck--1 telescope fitted with an aperture mask. The main aim of this paper is to provide unique observational data that could help to understand the nature of the mysterious 2.1--yr oscillation in radial velocity for this star.

\section[Observations and data reduction]{Observations and data reduction} 

\begin{table}
\centering
\caption{\label{calibrator_info} Calibrator information.}
\begin{minipage}{140mm}
\begin{tabular}{l c c r}
\hline
\hline
{Calibrator} & {Spectral} & {Adopted UD} & {Reference(s)}\\
{name} & {type} & {(mas\footnote{milliarcseconds})} & { }\\
\hline
{${\alpha}$ Lyr} & {A0V} & {3.22{$\pm$}0.01} & {\cite{2006A&A...452..237A}}\\
{${\beta}$Her} & {G7IIIa} & {3.40{$\pm$}0.03} & {This work}\\
{${\rho}$ Ser} & {K5III} & {3.28{$\pm$}0.04} & {\cite{2002A&A...393..183B}}\\
{${\alpha}$ Aql} & {A7V} & {3.46{$\pm$}0.04} & {\cite{2001ApJ...559.1155V}}\\
\hline
\end{tabular}
\end{minipage}
\end{table}

Observations were performed at the IOTA interferometer  and at the Keck--1 telescope. IOTA was a long--baseline optical interferometer located at the Smithsonian Institution's Whipple Observatory on Mount Hopkins, AZ. IOTA  operated from 1995--2006, and was used as  a testbed for new cutting--edge technologies \citep{2001A&A...376L..31B,2002SPIE.4838..1127}.  IOTA produced a large number of astronomy results over the past few years \citep{2002ApJ...579..446M, 2003A&A...408..553O, 2004ApJ...602L..57Mod, 2004A&A...418..675P, 2005AJ....130..246K, 2006ApJ...645L..77M, 2006ApJ...647..444M, 2007ApJ...659..626Z, 2007A&A...466..649K, 2008ApJ...679..746R, 2008arXiv0804.0192L}.

Observations performed at the Keck--1 telescope used the near infrared camera (NIRC) and an aperture mask that converted the telescope pupil into a sparse interferometric array of 9.8~m maximum baseline. For detailed discussion of the Keck aperture--mask experiment and scientific rationale see \cite{2000PASP..112..555T}.

In  Table~\ref{observation_log} we  present a journal of our observations,  listing date, filter and calibrator star.  In Table~\ref{calibrator_info} we detail the physical properties of the calibrators. Our  H--band  data  used the IONIC combiner \citep{2003SPIE.4838.1099B} with  narrow H--band  filters  at  the  IOTA interferometer and  at the Keck telescope  for the observations of 2004, while data  was acquired using  a standard  H--band filter  at IOTA  for 2005.  Data from 2006 used a low--dispersion spectrograph which provided seven spectral channels  across the H--band with an R=39.  For  a description of  the spectrograph see \cite{2003SPIE.4838.1225R} { and} \cite{2008SPIE.7013E..88P}. First results with the spectrograph were published by \cite{2008arXiv0804.0192L}.

{ For the  aperture--masking experiment} we refer the readers to the work of \cite{1999PhDT........19M} and \cite{2000PASP..112..555T} for the data--analysis, procedures adopted to extract visibilities and closure--phases in OIFITS format \citep{2005PASP..117.1255P}.
\begin{figure}
\includegraphics[scale=.38]{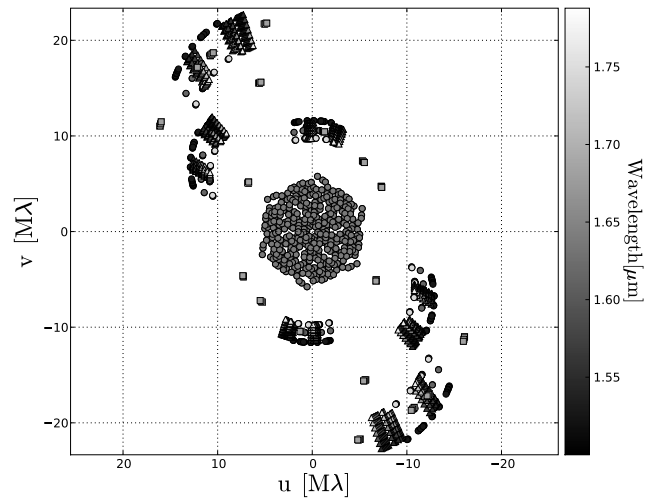}
\caption{\label{uv_coverage}The \uv coverage for CH Cyg for 2004 (circles) 2005 (squares) and 2006 (triangles). The shades of grey represent the wavelength of
the data points.}
\end{figure} 

The data reduction pipeline for the IONIC combiner was described in detail in \cite{2004ApJ...602L..57Mod}. Briefly, reduction of the squared visibilities (V\textsuperscript{2}) followed the same method explained by \cite{1997A&AS..121..379C}. Interferograms were corrected for intensity fluctuations and bias terms from readout noise and photon noise. The power spectrum of each interferogram was calculated in order to measure V\textsuperscript{2}. A transfer matrix was used to take in account the variable flux ratio for each baseline. The absolute calibration accuracy was studied by \cite{2004ApJ...602L..57Mod} by observing single stars of known size. 

Closure phases for the IONIC combiner were obtained using two independent methods; one was developed by \cite{1996A&A...306L..13B} for the Cambridge optical aperture synthesis array (COAST), and the other by \cite{2003SPIE.4838..387H} for the infrared spatial interferometer (ISI). In order to measure meaningful closure phase, fringes must at least be present in three baselines and the fringe packets must overlap, to be detected in the same coherence time.  The largest error in closure phase { offset for a point source} was caused by chromaticity in the combiner which limits the absolute precision when source and calibrator are not the same spectral type. Engineering tests performed by \cite{2004ApJ...602L..57Mod} showed that the closure phase varied systematically by 1.4 {$\pm$} 0.3\textsuperscript{o} between a cool star of spectral type M3 and a hot B8 star. 

The IOTA data pipeline produced visibility and closure phases in OIFITS format, which can be easily imported in imaging or modelling programmes using libraries provided by John Young, for C and Python\footnote{http://www.mrao.cam.ac.uk/research/OAS/oi\_data/oifits.html} and John Monnier for IDL\footnote{http://www.astro.lsa.umich.edu/~monnier/oi\_data/index.html}. A standard 2\% systematic error was added { in quadrature} to the visibility and closure--phase data as in \cite{2004ApJ...602L..57Mod}. Calibrated data in OIFITS format will be made available on request for interested investigators. 
 
{\section{Modelling}\label{model}} 
\cite{2004ApJ...604..800W} conducted a thorough review on the causes of the LSP variations in asymptotic giant branch (AGB) stars. They ruled out several possible models, among which were radial pulsation, companion in close orbit, spots on the star and modulation from an ellipsoidal--shaped AGB star. They  concluded that non--radial pulsation was the most likely explanation for LSP. In their recent paper \cite{2008arXiv0811.0631H} after a thorough review of the literature on CH~Cyg applied a similar approach to rule out possible models explaining the 2.1--yr change in radial velocity in CH~Cyg. We used our interferometry data to verify some of the hypotheses discussed in these papers. Due to  limited \uv--plane coverage of the data (see Figure~\ref{uv_coverage}), in particular for the 2005 epoch, we could not  resort to model--independent imaging of the CH Cyg system. For this reason we used parametric modelling to derive the size of the star, the FWHM size of the dust and the position and distance of the asymmetries detected in the closure--phase data. For model fitting we used publicly available least--squares minimisation routines\footnote{A non--linear least squares curve fitting (MPFIT), developed by Craig { Marquardt}. http://cow.physics.wisc.edu/craigm/idl/ }.

Our modelling was similar to \cite{2008ApJ...679..746R} except that we did not need to model multi--wavelength sizes for the star, since CH~Cyg does not change size appreciably with respect to  wavelength. We  decided to test the following hypothesis to { investigate} the cause of the LSP variations and to interpret our data: (1) radial pulsation of the star, (2) presence of dust inside the 15.6--yr orbit (3) spots on the star, (4) M~giant companion in a 15.6--yr orbit, (5) dwarf companion in a 2.1--yr orbit, (6) non--radial pulsation. \newline

\subsection{\label{radial_pulsation}Radial pulsation and dust}

\begin{figure}
\includegraphics[scale=.35,angle=90]{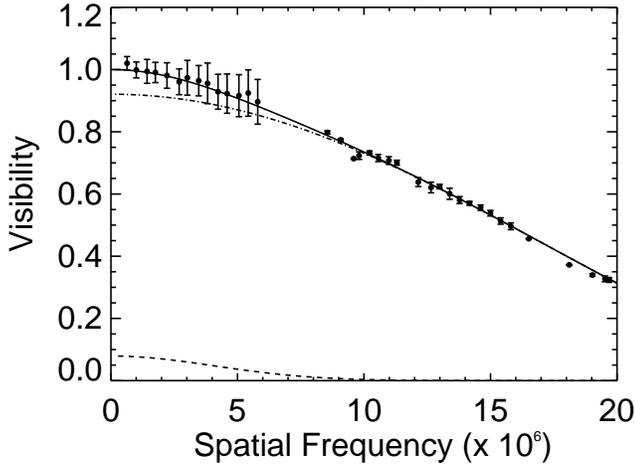}
\caption{\label{visibility_plot}Visibility plot comprising all data from 2004 to 2006. A simple uniform disc plus a Gaussian disc model representing the star and the dust are superposed to the data. The contributions from the dust (dashed line) and the star (dash-dotted line) are also over-plotted for clarity.}
\end{figure}

In order to test (1) and (2) a simple model composed of a uniform disc (UD) for the star and a Gaussian disc (GD) for the dust was first attempted in order to obtain a size for the star and for the dust.   All data from all epochs were used for this model since, by visual inspection, our visibility points superposed quite well, indicating that the size of the star did not change appreciably outside the error bars of the data, neither with time nor wavelength nor position angle. 

Figure~\ref{visibility_plot} shows the result of the fit. The data were smoothed using an azimuthal average due to the otherwise very large number of data points present on the graph. For each bin we used the mean of the original data points weighted by their errors. The error on each new data point was the standard deviation for the bin. The fit was performed on the original and non--smoothed data.
\begin{table}
\centering
\caption{\label{size_of_star} Size of star and size of the dust--shell emission. We used a uniform disc to model the star and  a Gaussian disc to model the dust. A model with a spherical dust--emission and a model with an elliptical dust--emission were attempted. The two models produced a very similar reduced $\chi^2$}
\begin{minipage}{140mm}
\begin{tabular}{l c c c c c r}
\hline
\hline
{Size}       & {FWHM} &{Flux ratio}     &{P.A.} &{Axis ratio}&{ Reduced}\\
{M~giant} & {dust\footnote{Size of the dust emission.}} &{(M~giant}     &{    } &{dust}&{${\chi^2}$}\\
{(mas)}      & {(mas)}&{/dust)}&{({\deg})}&{(M/m)}&{          }\\
\hline
{8.74{$\pm$}0.02}     &{19.1{$\pm$}1.0}&{11.6{$\pm$}0.3}&{0.0}&{1.0}&{1.3}\\
{8.74{$\pm$}0.02}&{19.2{$\pm$}0.9}&{11.6{$\pm$}0.1}&{103{$\pm$}5}&{1.28{$\pm$}0.01}&{1.2} \\
\hline
\end{tabular}
\end{minipage}
\end{table}
Table~\ref{size_of_star} shows the parameters obtained from the fit. The value of  8.74{$\pm$}0.02~milliarcseconds (from now on mas), for the diameter of the M~giant is the most accurate so far thanks to the large amount of data used. This value is close to the value of 7.8{$\pm$}0.6 obtained with infrared interferometry in June~2001 by \cite{2003SPIE.4838.1043H} using a simple UD fit. The errors were derived using bootstrap statistics on the data set. The full--width half maximum size of the Gaussian dust emission was 19.13{$\pm$}1.00~mas or 2.2{$\pm$}0.1 stellar { diameters} FWHM, showing that hot dust exists close to the M~giant.  A marginally improved reduced ${\chi^2}$ was obtained by fitting an elliptical dust distribution around the star. However the difference in reduced ${\chi^2}$ was too small in order to justify an asymmetric model for the dust emission in the near infrared. The parameters from the elliptical dust--emission model are also listed in Table~\ref{size_of_star}.

\cite{2002ApJ...577..447T} monitored an oxygen--rich and a carbon--rich Mira star measuring the change of angular size with respect to the  pulsation cycle at the Palomar testbed interferometer (PTI). We did not detect any such change in CH~Cyg. Unfortunately our coverage of the 2.1--yr period was quite limited (basically two points at phase 0.4 and one point at phase 0.9 of the ``orbital'' period). This coverage is insufficient to completely rule out radial pulsation for this star. However, we notice that \cite{2003SPIE.4838.1043H} obtained a diameter of 7.8$\pm$0.6 mas in June 2001, using a simple UD model with three visibility points. Considering the crude UD model used that does not take in consideration the dust shell, this diameter is not very different from our measurement and would indicate that the star did not change diameter with time. Also, radial pulsation was ruled--out by \cite{2008arXiv0811.0631H} as the cause for secondary period in CH~Cyg since the period--luminosity relation for AGB stars \citep{1990AJ.....99..784H} would produce a period of about 250 days for a K=-7.5 star not 770 days. 

\subsection{\label{spots_on_star}Spots on the star}
In order to model the closure--phase signal expected from a spotted star, we used an additional uniform disc that could be placed at different position angles and separation from the centre of the UD~+~GD model representing the M~giant and the dust emission. Flux ratios between the M~giant and the companion/spot  and between the M~giant and the dust  were allowed to change.  The size of the additional UD was also free to change. The three epochs  were analysed separately in order to detect asymmetries that would change with the observing epoch. We performed a parameter--space search in { an} attempt to identify the position of the asymmetry.

We could not find any solution with an unresolved or moderately resolved spot on the surface of the star. All the solutions converged  to structure outside the disc of the M~giant unless we restricted the flux to 10\% or more of the flux of the M~giant as done in Section~\ref{Mgiant_companion}. 
\subsection{\label{Mgiant_companion}M~giant companion on the 15.6--yr orbit} 
We tested the hypothesis that the companion is a red giant in the 15.6--yr orbit as discussed in the model proposed by \cite{1996A&A...308L...9S}. That model was devised to explain the eclipses observed in the 15.6--yr orbit and kept the symbiotic pair in the 2.1--yr orbit, given that there are no known symbiotic stars found on an orbit of period as long as 15.6~yrs.  We restricted the flux ratio of the M~giant/companion to values around  8.6, as expected in \cite{2004ARep...48..813T}.  The field of view (FOV) of IOTA was limited by the bandwidth of the photometric filter used:  $\mathit{FOV}=\lambda ^{2}/\Delta \lambda B$, where ${\lambda}$ is the wavelength $\Delta \lambda$  the bandwidth and B the baseline. For the largest bandwidth used (0.3 {${\mu}$}m at 1.65 \ {${\mu}$}m) and a baseline of 38 metres we obtained a minimum FOV of 50 mas. We performed a 50 mas wide search in all our data sets. 

We did not find any trace of a companion in our best data sets of 2004 and 2006 when using the Taranova flux { ratio.} A second red giant should have been evident in the data. In particular the Keck telescope aperture--masking experiment should have easily detected a second giant star down to a flux ratio M~giant/companion of about 100 { \citep{2008ApJ...678..463I,2008ApJ...679..762K}}.
\subsection[The orbit of CH Cyg]{Dwarf companion in a 2.1--yr orbit} 
\begin{figure*}
\includegraphics[scale=.95]{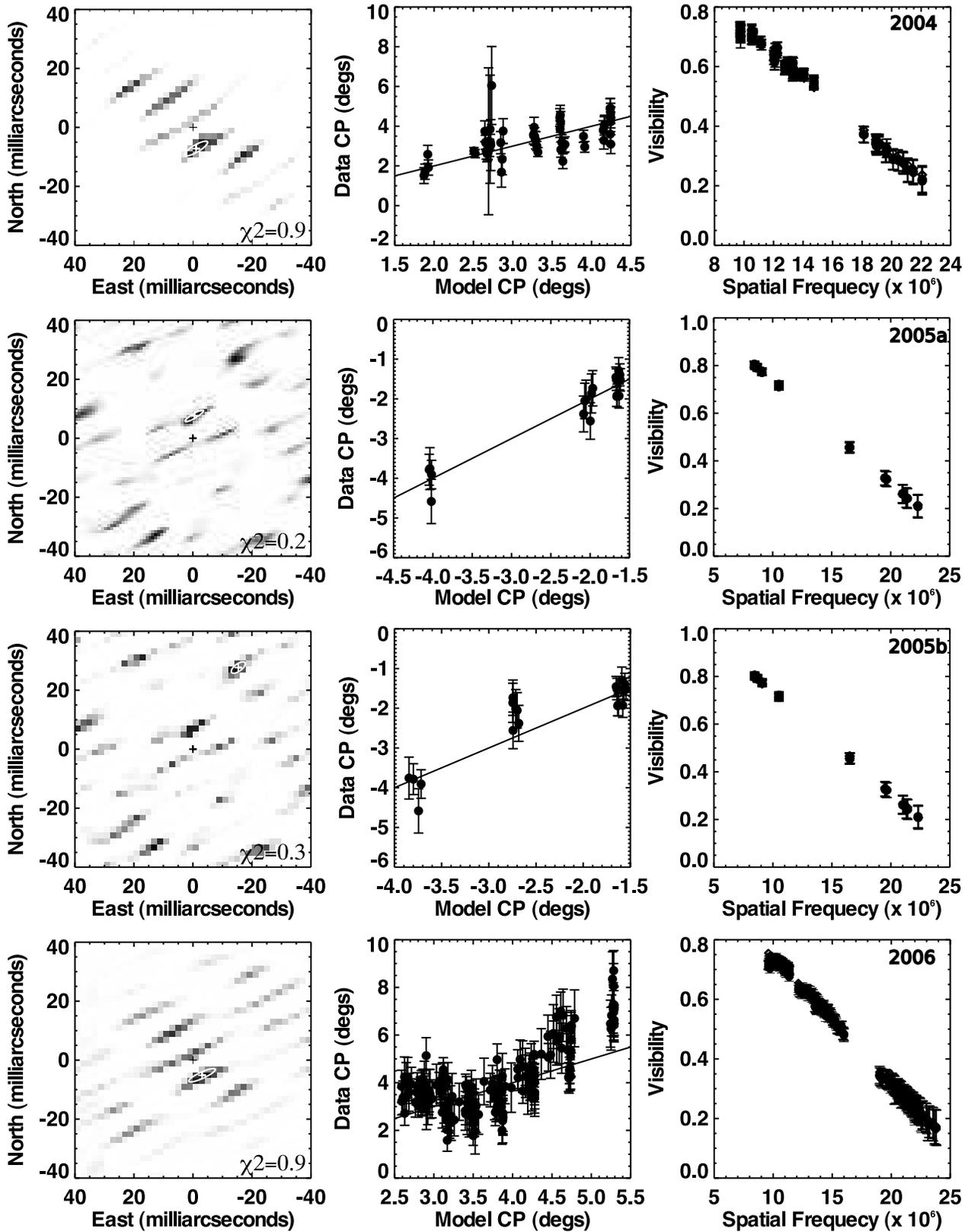}
\caption{\label{likelihood_surface}Binary--star search in $\chi^2$ space. { A simple binary-star model was fitted to the visibility and closure--phase data from different epochs}. The M~giant was kept at the centre of the field while the companion was placed in all possible positions of a $40 \times 40$ mas square grid. A reduced $\chi^2$ value was obtained for each position and the values recorded on a two-dimensional array.  Left column shows a likelihood surfaces derived from the  $\chi^2$ arrays for  our 2004, 2005a and 2005b (non--unique solutions) and 2006 data. The positions of the asymmetries are encoded in the likelihood map. The white crosses represent the positions of the asymmetry and the white ellipses encode the uncertainty of the position. The centre column shows data--versus--model plots for closure phase. The right column shows visibility--versus--spatial--frequency plots (filled circles) with superposed points derived from the model (open diamonds). Note that the high density of data points makes it very hard to distinguish between filled circles and open diamonds. We observed that the closure phase flipped sign between 2004 and 2005 and between 2005 and 2006 meaning that the the detected asymmetry was in the opposite direction in 2005.}
\end{figure*}

We tested the hypothesis of a faint companion orbiting the M~giant as in Section~\ref{spots_on_star} and Section~\ref{Mgiant_companion}. We restricted the flux contribution of the companion to less than 2\% of the total flux in order to simulate a large $\Delta$m between the M~giant and the companion. As a consequence we found asymmetries outside the M star in all three data sets. Figure~\ref{likelihood_surface} shows the likelihood maps obtained from the reduced ${\chi^2}$ surfaces. The dotted--line ellipses are the errors on the positions of the companion which are quite large in the East--West direction due to the limited \uv coverage of the IOTA in that direction. 
\begin{table}
\centering
\caption{\label{orbital_pos} Orbital positions for the low--mass object.}
\begin{minipage}{140mm}
\begin{tabular}{l c c c c c c r}
\hline
\hline
{Mean}      & {Sep}   & {P.A}                      & {Flux ratio}       & {Flux ratio} & { Reduced}\\
{JD}        & {(mas)} & {(\textrm{{\deg}})} & {M~giant/}        & {M~giant/}  & {${\chi^2}$}\\
            &         &                            & {dwarf}            & {dust}       & { }\\
\hline
{2453122.9} & {7 \textsupsub{+3}{-2}} & {188 \textsupsub{{}+37}{-26}} & {78{$\pm$}1} & {9.6{$\pm$}0.4} & {0.9}\\
{2453529.0} & {8\textsupsub{+3}{{}-2}} & {356\textsupsub{{}+4}{-4}} & {88{$\pm$}5} & {18.7{$\pm$}1.0} & {0.2}\\
{2453529.0} & {32\textsupsub{+3}{{}-3}} & {331\textsupsub{{}+3}{-3}} & {104{$\pm$}4} & {18.4{$\pm$}0.8} & {0.3}\\
{2453855.2} & {6\textsupsub{+2}{{}-1}} & {211\textsupsub{{}+37}{-42}} & {74{$\pm$}1} & {14.2{$\pm$}0.4} & {0.9}\\
\hline
\end{tabular}
\end{minipage}
\end{table}

Table~\ref{orbital_pos} lists the  separations, position angles, flux ratios and reduced ${\chi^2}$ of the asymmetries for the three epochs. The UD size of the companion converged to a point source for all epochs. The error bars on the parameters were derived from the error ellipses. The 2005 data set converged to two separate solutions: one at 8~mas separation and another 32~mas separation. The second solution  was likely due to a degeneracy caused by the limited amount of data available for the 2005 epoch. We could fit the 2.1--yr elliptical orbit to the 32~mas position, with a ${\chi^2}$ of 0.3. The semi-major axis of this orbit was 25.6{$\pm$}0.8 mas which produced a far { smaller} luminosity and a much shorter distance than expected for this star. For this reason we excluded this solution.  We must point out that \cite{2007AstBu..62..339B} detected a faint companion  with speckle interferometry in 2004 at 43{$\pm$}1~mas separation  and 24.1\deg{$\pm$}2.1\deg  position angle. However we do not believe that our 32~mas position is related to this detection. In fact the Keck telescope aperture--masking experiment should have easily detected a companion down to a flux ratio M~giant/companion of about 100 in our 2004 data { \citep{2008ApJ...678..463I,2008ApJ...679..762K}} but such detection did not happen.

\begin{figure*}
\includegraphics[scale=1.26]{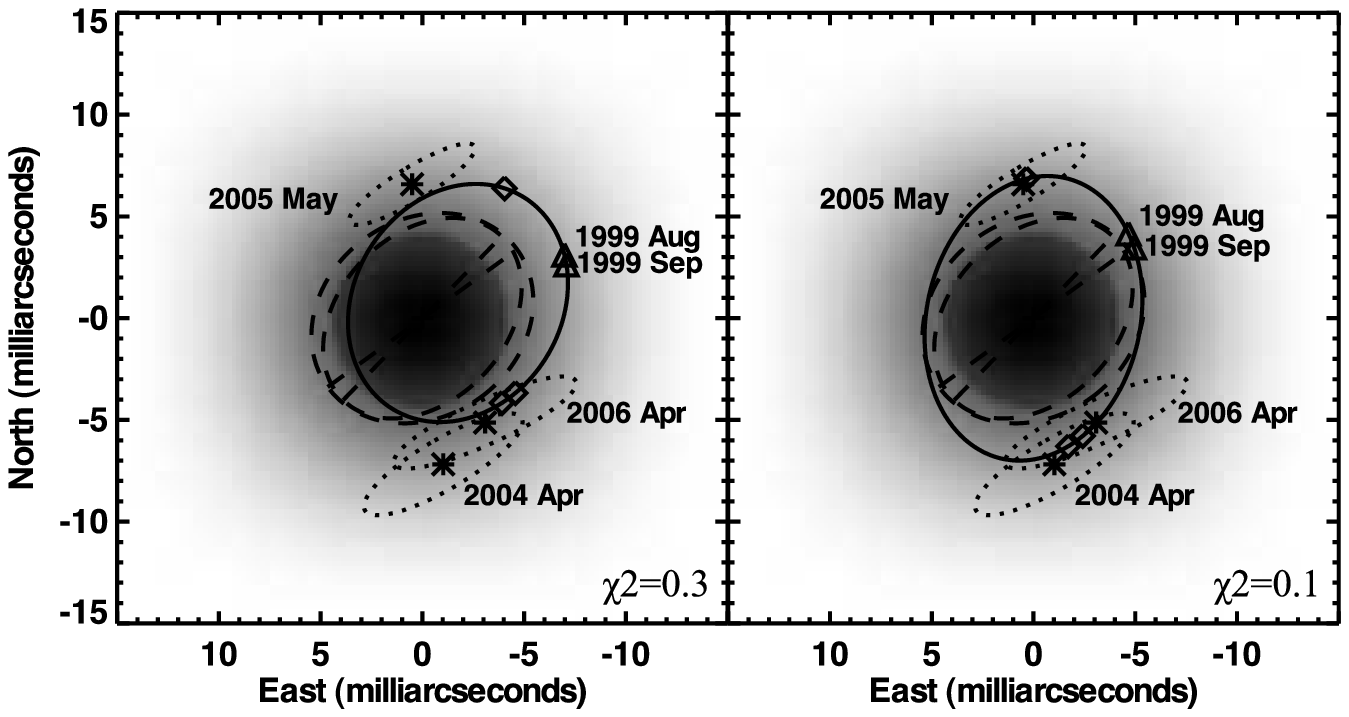}
\caption{\label{astrometric_orbit} The astrometric orbit of CH Cyg. The plots show an elliptical orbit fit (left) and a circular orbit fit (right). Superposed to the orbits are a uniform disc representing the star and a Gaussian disc representing the dust emission in the system. The flux contribution from the dust was exaggerated to render the dust extent visible in the picture. The diamonds are the expected positions of the companion relative to the M~giant, according to the ephemeris. The observed positions of the secondary component are marked with error ellipses (dotted line) centred around a star symbol. The triangles are the expected positions of the companion during the observations at the COAST interferometer. The dashed--line ellipses represent an elliptical model of CH~Cyg from Young et al. (2000). The major axis of the ellipses is also shown to better appreciate the orientation of the ellipses.}
\end{figure*}

In order to { investigate} the { hypotheses} that the detected asymmetries were the signature of a faint companion in a 2.1--yr orbit we attempted  orbit fits to the astrometric positions derived from the IOTA closure--phase data using infrared radial--velocity  orbital solutions  from \cite{2008arXiv0811.0631H}.  We attempted orbital fits using a circular orbit \citep{2008F} which was discarded in \cite{2008arXiv0811.0631H} due to the large residuals in the orbit fit. We also used the elliptical orbit solution from \cite{2008arXiv0811.0631H}. We obtained a reduced {${\chi^2}$} of 0.1 for the circular orbit and a  reduced $\chi^2$ of 0.3 for the elliptical orbit. 
{ Such small reduced ${\chi}^{2}$ values are possible given the small number of degrees of freedom (3 from the 6 data points and 3 free parameters) and the quite large error bars on the astrometric positions}. The obtained orbits are shown in Figure~\ref{astrometric_orbit}. The combined orbital parameters from radial velocity and interferometry and some derived parameters are shown in Table~\ref{orbital_par}. The first part of the table lists the orbital parameters from \cite{2008arXiv0811.0631H} and \cite{2008F}.  The errors on the parameters obtained from interferometry were derived using  { Monte Carlo} simulations.   
\subsection{Non--radial pulsation} 
Although we could not obtain acceptable reduced $\chi^2$ from the ``spots on the star'' model we managed to obtain reasonably good fits in simulations of large asymmetries on the star. We used low--order spherical harmonics to  simulate  large flux variations  (up to 20\%) across the star. Such a dramatic brightness change could simulate the closure phase signal of CH Cyg but seemed an unlikely explanation even in term of non radial pulsation: we are not aware of any physical mechanism that could produce such a dramatic  change of brightness across a star. Figure~\ref{non_radial} shows the models and the corresponding fits of visibility and closure phase. The reduced ${\chi^2}$ was reasonably close to the dwarf--companion model. The asymmetry appears to rotate with the 2.1--yr period.

\section{Discussion}
\cite{2008arXiv0811.0631H} restricted the possible explanation for the 2.1--yr secondary period of CH~Cyg to a low--order g--mode non--radial pulsation of the M~giant or a to low--mass companion (0.2 M\sun) in close orbit to the M~giant.  In this model the companion responsible for the activity is on the 15.6--yr  orbit.

According to \cite{2008arXiv0811.0631H} a 0.2~M\sun companion would have a temperature of about 3200K and would be spectroscopically indistinguishable from the M~giant. Since CH~Cyg is  single--lined binary/triple star the masses of the components cannot be derived directly. We can however test the derived parameters against the published literature, assuming the mass of one of the components. Table~\ref{derived_par} shows the change of the derived parameters for different values of the mass of the companion. M$_1$ is the mass of the M~giant in solar masses, R$_1$ the radius  in solar radii,  L$_1$ the luminosity in solar luminosities. The semi--major axis ``$a$'' of the orbit in physical units of AU was obtained using Kepler's law. D is the distance in parsecs. The table is divided in two parts, one concerned with the circular orbital solution and one with the elliptical solution. 
\begin{table}
\centering
\caption{\label{orbital_par} { Orbital parameters of the possible low--mass object.}}
\begin{minipage}{140mm}
\begin{tabular}{l c c c r}
\hline
\hline
\textbf{Parameters}&\textbf{Circular}& &\textbf{Elliptical}\\
\textbf{          }&\textbf{solution}& &\textbf{solution}\\
\hline \\
\multicolumn{4}{c}{\textbf{Radial velocity}} \\
\\
\textbf{ \textit{P} (days)}&{749.8{$\pm$}2.3}& &{750.1{$\pm$}1.3}\\
\textbf{ \textit{T0} (HJD)}&{2446823.2{$\pm$}7.7}& &{2447293.5{$\pm$}12.9}\\
\textbf{ ${\omega}$ ({\textordmasculine})}&{0.0}& &{229.5{$\pm$}7.7}\\
\textbf{\textit{e}}&{0.0}& &{0.330{$\pm$}0.041}\\
\textbf{\textit{K} (Km s\textsuperscript{{}-1})}&{2.87{$\pm$}0.13}& &{2.87{$\pm$}0.13}\\
\textbf{\textit{${\gamma}$} (Km s\textsuperscript{{}-1})}&{{}-59.93{$\pm$}0.10}& &{{}-59.91{$\pm$}0.09}\\
\textbf{\textit{a sin i} (Km)}&{2.96x10\textsuperscript{7}{$\pm$}0.29x10\textsuperscript{7}}& &{2.79x10\textsuperscript{7}{$\pm$}1.23x10\textsuperscript{7}}\\
\textbf{\textit{f(m) }(M\sun)}&{0.00018{$\pm$}0.0002}& &{0.00015{$\pm$}0.0002}\\
\\
\multicolumn{4}{c}{\textbf{Interferometry}} \\
\\
\textbf{ \textit{i} ({\textordmasculine})}&{138{$\pm$}10}&  &{146{$\pm$}6}\\
\textbf{ ${\Omega}$ ({\textordmasculine})}&{347{$\pm$}7}& &{337{$\pm$}8}\\
\textbf{ \textit{a} \ (mas)}&{7.1{$\pm$}0.3}& &{6.3{$\pm$}0.3}\\
\hline
\end{tabular}
\end{minipage}
\end{table}
 
The mass of 2~M\sun from \cite{2008arXiv0811.0631H}, a luminosity of 6900~L\sun from \cite{2006ApJ...647..464B} and a radius of 280$\pm$65~R\sun obtained by \cite{1999ESASP.427..397S} through infrared spectroscopy were used to restrict the solutions listed in Table~\ref{derived_par}. A value of 0.32~M\sun for the low--mass companion yielded a mass of 2~M\sun  a radius of 250~R\sun and a luminosity of 6517 L\sun for the M~giant, very close to the values from the literature.  

Also, the distance obtained from the size of the circular orbit derived using Kepler's law and the apparent size of the orbit in milliarcseconds, was 296~pc, comparable within errors to the 244{\textsupsub{+49}{-35}}~pc distance obtained from the revisited data reduction of the Hipparcos parallax \citep{2007ASSL..250.....V}.

\cite{2004AcA....54..347S} found that Wood's sequence--D variables are a continuation of sequence--E ellipsoidal variables.  \cite{2007AcA....57..201S} require that the ratio between the radius of the star and the semi--major axis of the orbit should be $R/a \sim 0.4$ for the binary explanation of the LSP (Equation~5 in \cite{2007AcA....57..201S}). For our circular orbit solution the ratio derived by the angular diameter of the star and the semi--major axis of the hypothetical orbit is 0.6, \abt50\% larger than the required value.
\begin{figure*}
\includegraphics[scale=.9]{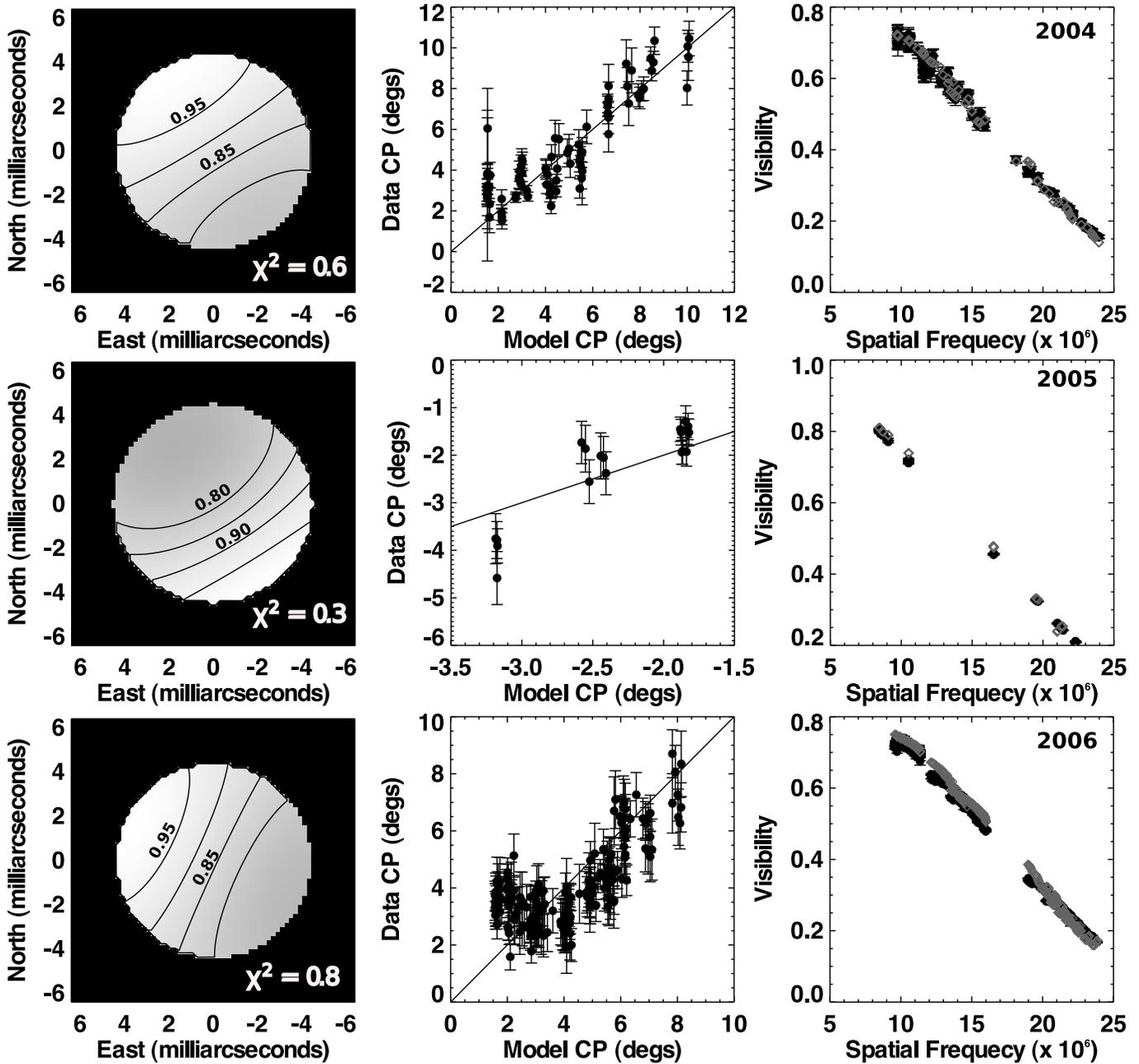}
\caption{\label{non_radial} The 2004 to 2006 visibility and closure--phase data was also fit by an asymmetric brightness distribution on the surface of the star. The left column shows models of the star for different epochs. The asymmetric flux distribution was modelled using spherical harmonics. The centre column shows data--versus--model plots for the closure phase. The right column shows visibility--versus--spatial--frequency plots (filled circles) with superposed points derived from the model (open diamonds). The asymmetric change in brightness across the star could be caused by a non-radial pulsation \citep{2004ApJ...604..800W,2008arXiv0811.0631H}. However we are not aware of any pulsation  mechanism that would produce a 20\% change in brightness across the star. { The closure phase changed sign and the }asymmetry flipped of 180\deg in 2005.}
\end{figure*}

\cite{2007ApJ...660.1486S} also proposed a model { where the LSP variation are} caused by mass loss from the giant to the low--mass companion.  Since we detected hot--dust emission inside the possible 2.1--yr orbit we cannot exclude that LSP photometric variation are caused by dust trailing the low--mass companion.  There have been claims of eclipses in the 2.1--yr orbit of CH~Cyg \citep{1996A&A...308L...9S,1998MNRAS.297...77I}. The inclination of the circular orbit obtained from the interferometric data would prevent eclipses but if the dust is clumpy and is trailing the companion it could be responsible for occasional photometric variations and could simulate eclipses.
\begin{table}
\centering
\caption{\label{derived_par} { Derived red giant parameters as a function of the faint companion's mass.}}
\begin{minipage}{140mm}
\begin{tabular}{l c c c c c c r}
\hline
\textbf{ M$_2$ (M\sun) }&{0.1}&{0.2}&{0.3}&{0.4}&{0.5}&{0.6} \\
\\
\multicolumn{7}{c}{\textbf{Circular orbit}} \\
\\
\textbf{ M$_1$ (M\sun) }&0.3 & 0.9 & 1.8 & 2.8 & 3.9 & 5.2 \\
\textbf{ R$_1$ (R\sun) }&156  & 221 & 270 & 312 & 349 & 382 \\
\textbf{ L$_1$ (L\sun) }&2037 &4073&6110&8146&10183&12219\\
\textbf{ a (AU) }& 1.2 & 1.7 & 2.0 & 2.3 & 2.6 & 2.9  \\
\textbf{ D (pc) }&166& 235 & 288 & 332 & 371 & 407  \\
\\
\multicolumn{7}{c}{\textbf{Elliptical orbit}} \\
\\
\textbf{ M$_1$ (M\sun) }&0.2 & 0.7 & 1.4 & 2.2 & 3.2 & 4.2 \\
\textbf{ R$_1$ (R\sun) }&165  & 233 & 285 & 330 & 369 & 404  \\
\textbf{ L$_1$ (L\sun) }&2267&4534&6801&9068&11335&13601\\
\textbf{ a (AU) }& 1.5 & 1.6 & 1.9 & 2.2 & 2.5 & 2.7  \\
\textbf{ D (pc) }& 143 & 248 & 304 & 351 & 392 & 430 \\
\hline
\end{tabular}
\end{minipage}
\end{table} 

Another clue in favour of the low--mass companion explanation for  the LSP variation of CH~Cyg comes from the independent interferometric observations of \cite{2000SPIE.4006..472Y} at the wavelength of 905~nm. The visibility and closure--phase data from the COAST interferometer were modelled by an elliptical, limb--darkened star (parameters from that model are reproduced in Table~\ref{COAST_model}). Figure~\ref{astrometric_orbit} shows the astrometric orbit of CH~Cyg for the elliptical and circular orbit solution superposed to the models from \cite{2000SPIE.4006..472Y}. Interestingly the minor axis of the two ellipses is very close to the radius of the M~giant obtained from our model, while the major axis intersects the predicted orbital position of the companion on the circular orbit. The major axis of the ellipses did not intersect the orbital positions of the companion on the elliptical orbit. Also, our elliptical orbit fit to  our astrometric positions had a $\chi^2$ 3--times worse than the circular orbit fit. 
\begin{table}
\centering
\caption{\label{COAST_model} Elliptical star model parameters from Young et al. (2000).}
\begin{tabular}{l c c r}
\hline
\hline
{Epoch} & {Major axis} & Axial ratio & {P.A}  \\
        & {(mas)}      &             & {(\textrm{{\deg}})}\\
\hline
{99/08} & {11.5 $\pm${0.2}} & {0.84 $\pm${0.05}} & {126{$\pm$}9}\\
{99/09} & {11.2 $\pm${0.2}} & {0.79 $\pm${0.03}} & {136{$\pm$}5} \\
\hline
\end{tabular}
\end{table}

According to \cite{2008arXiv0811.0631H} the most powerful argument against the close--binary explanation of LSP variation in CH Cyg is the shape of the radial--velocity curve: the most likely orbit for the low--mass companion would be elliptical due to the large radial--velocity residuals obtained from fitting a circular orbit.  On the other hand \cite{2008arXiv0811.0631H} derived a mass of 2.0~M\sun for the M star, based on evolutionary arguments and argued that the Roche lobe for a 2~M\sun--~0.2~M\sun binary would constantly change from $(1+e)a$ at apoastron to $(1-e)a$ at periastron for an elliptical orbit. A 280~R\sun giant would fill the Roche lobe at each periastron passage generating large mass loss.  We argue that distortion by proximity effect \citep{2008ApJ...681..562E} could also change the shape of the radial velocity curve and therefore the circular orbit solution cannot be eliminated on the basis of this argument. 

\cite{2008arXiv0811.0631H} argue that non--radial pulsation could also reproduce the observed radial velocities and photometric variation of CH~Cyg. The problem with the non radial pulsation argument is that low--order g modes are evanescent in convective regions and there is no known physical mechanism that could explain the non--radial pulsation for  M~giant stars where radiative transfer is  mostly convective. As we show in Figure~\ref{non_radial} an asymmetric brightness distribution on the surface of the star could also reproduce our observed closure--phase signature. The flux variation across the star must be very large (20\%) in order to explain the closure--phase results and we are not aware of any physical mechanism that could produce such a dramatic  change of brightness across a star.

Close encounter with another object can produce non--radial oscillations on a fluid star through tides according to \cite{1990A&A...229..457E}. Circularisation of early--type main sequence binaries are also known to cause non--radial g--mode resonant oscillations. In late type stars turbulence is very efficient in { damping} these oscillation and circularising binary orbits.  \cite{2004MNRAS.353.1161I} studied tides in fully convective stars. They came to the conclusion that resonant tides may be possible in fully convective stars.

CH~Cyg is a very complex object. If it is a triple system the interactions of the companions with the M~star could be very complex. The signature in the radial velocity  could be caused by a combination of movement of the low--mass companion and non--radial pulsation of the star. This could explain the residuals found in fitting a circular orbit to the radial--velocity data. The non--radial pulsation may be due to tidal interaction of the low--mass companion with the M~giant. Such interaction would also cause rapid circularisation of the orbit for the low--mass companion. 

It is not clear what the timescale for the circularisation of the orbit and the dissipation of the non--radial pulsation would be. If the dissipation of the non--radial pulsation by convective turbulence is efficient and the timescale for circularisation short it is hard to explain why at least 25\% of stars in globular clusters show LSP variations. We suggest that in globular clusters interactions with low--mass companions could be more frequent than expected. LSP variations then would be caused by non--radial pulsations excited by orbital capture of a companion or circularisation of an elliptical orbit.

\section[Conclusions]{Conclusions}  
We have presented simple  models in order to explain  the  asymmetries detected  through infrared interferometry in the S--type symbiotic star CH  Cyg.  We  do  not   detect  significant  change  of  angular  size (8.74{$\pm$}0.02  mas) for  the M~giant  over a  3 year period, rendering radial  pulsation a less likely explanation for the 2.1--yr variability  in  radial  velocity   data.  We  find  a  spherical  hot dust--shell with an emission size  of { 2.2$\pm$0.1~D$_*$}~FWHM around the M~giant  star which could be responsible for some of the reported short--period eclipses. We find correlation between the 2.1--yr variability and the variation in our closure phase. We model the closure phase as a large change in brightness across the M~giant and/or a low mass companion in close orbit around the star. While the most likely explanation for the change in brightness can be a non--radial pulsation we argue that a low--mass companion in close orbit could be the physical cause of the pulsation. The combined effect of pulsation and low--mass companion in close orbit to the M~giant could explain the behaviour of the radial--velocity curves and the asymmetries detected in the closure--phase data. If CH~Cyg is a typical long secondary period variable then LSP variations could be explained by the effect of  an orbiting low--mass companion on the primary star.
\section*{Acknowledgments}
We acknowledge Leslie Hebb for a useful discussion. We thank Kenneth Hinkle and Francis Fekel for making available to us their data and results prior to publication and for many useful discussions and advice.  This research  was made possible  thanks to  a  Michelson Postdoctoral Fellowship  and  a Scottish  Universities  Physics Alliance  (SUPA) advanced fellowship  awarded to E.   Pedretti. N. Thureau received research funding from the European Community's Sixth Framework Programme through an International Outgoing Marie-Curie fellowship OIF - 002990.\\The IONIC project  is a collaboration  among  the   Laboratoire  d'Astrophysique  de  Grenoble (LAOG), Laboratoire d'Electromagnetisme Microondes et Optoelectronique (LEMO),  and  also CEA--LETI  and  IMEP,  Grenoble,  France. The  IONIC project  is funded  in  France  by the  Centre  National de  Recherche Scientifique and Centre National d'Etudes Spatiales. This research has made use of NASA's Astrophysics Data System Bibliographic Services and of the SIMBAD.  database operated at CDS, Strasbourg, France. 
\bibliographystyle{mn2e}
\bibliography{/Users/ep41/Documents/bib/astro,/Users/ep41/Documents/bib/detectors,/Users/ep41/Documents/bib/interf}

\end{document}